\documentclass[twocolumn,showpacs,preprintnumbers,amsmath,eqsecnum,amssymb]{revtex4}
\usepackage{graphicx}
\begin{document}
\preprint{IMAFF-RCA-03-06}
\title{Wormholes and ringholes in a dark-energy universe}
\author{Pedro F. Gonz\'{a}lez-D\'{\i}az}
\affiliation{Centro de F\'{\i}sica ``Miguel A. Catal\'{a}n'', Instituto de
Matem\'{a}ticas y F\'{\i}sica Fundamental,\\ Consejo Superior de
Investigaciones Cient\'{\i}ficas, Serrano 121, 28006 Madrid (SPAIN).}
\date{\today}
\begin{abstract}
The effects that the present accelerating expansion of the
universe has on the size and shape of Lorentzian wormholes and
ringholes are considered. It is shown that, quite similarly to how
it occurs for inflating wormholes, relative to the initial
embedding-space coordinate system, whereas the shape of the
considered holes is always preserved with time, their size is
driven by the expansion to increase by a factor which is
proportional to the scale factor of the universe. In the case that
dark energy is phantom energy, which is not excluded by present
constraints on the dark-energy equation of state, that size
increase with time becomes quite more remarkable, and a rather
speculative scenario is here presented where the big rip can be
circumvented by future advanced civilizations by utilizing
sufficiently grown up wormholes and ringholes as time machines
that shortcut the big-rip singularity.
\end{abstract}


\maketitle

\section{Introduction} Solutions to Einstein equations that
describe closed timelike curves (CTCs) and time machines are known
since a few years after the discovery of general relativity [1].
At present, apart of a number of interesting recent proposals that
have been also much debated [2], essentially three general kinds
of such solutions can be recognized. First of all, we have the
so-called wormholes [3] and their topological generalizations,
ringholes [4] and Klein bottle holes [5]. These correspond to
shortcuts of the structure of space-time itself which are based on
topologically generalizing Misner solution [6] for different
matter contents that show what is known as exoticity; i.e. the
violation of some classical energy conditions-particularly the
dominant energy condition [3,4,7]. By far, these are the most
popular and studied solutions. Three different topologies have
been so far considered for the holes: spherical [3], toroidal [4]
and that corresponding to a Klein bottle [5]. The second general
kind of spacetime containing CTCs was first introduced by Gott [8]
and corresponds to solutions to gravity with infinitely large
parallel cosmic strings which move with high velocity relative to
each other. These solutions do not entail any kind of exotic
matter, but lack physical reality because they cannot be fitted in
a finite universe. Finally, Gott and Li have considered [9] the
case for the creation of the universe from itself in a sea
-presumably the spacetime foam- of nontrivial topological
submicroscopic structures where there are CTCs. In particular, a
multiply connected de Sitter universe was considered. This was
proved to be classically stable and, if the regions containing the
CTCs were of the Planck size, quantum-mechanically stable as well
[10]. In this paper we shall restrict ourselves to consider
Lorentzian spacetime tunnelings which involve just Morris-Thorne
wormholes and ringholes.

Semiclassical calculations of the renormalized stress-energy
tensor in spacetimes containing CTCs (or more precisely,
satisfying the Misner symmetry) generally show a quantum
instability at the so-called chronology horizon (i.e. the Cauchy
horizon which is the onset of the nonchronal region filled with
CTCs), where that tensor diverges [11]. This fact was taken as a
basis by Hawking to introduce a much debated generalizing
conjecture for chronology protection, stating that the laws of
physics prevent CTCs to occur [12]. It was later argued [13] that
while such a conjecture could well apply to macroscopic closed
curves, by no means it can be applied to CTCs at the Planck scale;
i.e. typically the scale of the quantum spacetime foam. It is now
generally recognized that the chronology protection conjecture is
just a semiclassical proposal, with no implementation in any
spacetime with quantum structure. It would follow that quantum
CTCs and time travels are stable quantum-mechanically, and are
pervading the entire spacetime of the vacuum quantum foam. These
tentative conclusions will be taken for guarantee in this paper
where we shall investigate the way in which the size of stable
wormholes and ringholes, originally at the Planck scale in the
spacetime foam, is increased in an accelerating universe, while
their initial form is preserved. A similar study was done for
inflating wormholes by Roman some years ago [14]. The present
research may entails additional interest as: (i) it refers to
wormholes which are being growth {\it now}, and (ii) the rate at
which the scale factor increases might go beyond the exponential
blow up predicted by the presence of a cosmological constant if
the equation of state of the universe is characterized by a
parameter $\omega <-1$, a case which still is by no means excluded
by the latest observational constraints.

The proper context for the current placement of wormholes and
ringholes is that of an accelerating universe, nearly the seventy
percent of whose energy is in the form of what is known as dark
energy -the stuff making up the anti-gravitational vacuum of the
universe [15]. Latest observations and theoretical developments
[16] seem to favor the idea that dark energy might be in the form
of either a positive cosmological constant -the first and perhaps
yet the favorite explanation for current anti-gravity dominance-
or a slowly-varying quintessence scalar field or generalized
Chaplygin gas [17], equipped with a perfect-fluid equation of
state, $p=\omega\rho$, where $-.8\geq\omega\geq-1.5$. Actually the
most reliable constraint on $\omega$ has been provided by WMAP and
is given by $\omega<-0.78$ [18]. Note that a cosmological constant
corresponds to the case $\omega=-1$. Apart of the rather special
case of a finite and tiny time-independent cosmological constant
[19], up to now there are two observationally allowed regimes for
dark energy which are essentially different. If $\omega<-1$ the
energy density of the dark stuff will keep decreasing with time
and can induce an ever accelerating universe of the kind we are
able to observe. It has been pointed out that this will inexorably
induce the emergence of a future event horizon, so preventing the
formulation of any consistent fundamental theory based on defining
an S matrix [20]. Even more bizarre are the predictions arising
from the regime $\omega<-1$. In this case, there is a violation of
the dominant energy condition [21], that is $p+\rho<0$, that leads
to the fact that the energy density increases with time up to an
infinite value, at which point the size of the universe blows up
at a finite time. This is known as big rip and represented the end
of the universe and everything in it as induced by a kind of dark
energy which is denoted as phantom energy [22]. Of course the
problem of future event horizon in this regime is not only still
present but it is made even more acute.

Since the dominant energy condition is violated in the regime with
$\omega<-1$, the possibility is opened [7] in that regime for the
occurrence of the above discussed wormholes and ringholes. It
appears then interesting to investigate the time evolution of such
holes when they are embedded in an accelerating universe endowed
with dark energy. In particular, we shall study in this paper the
way in which the size and shape of wormholes and ringholes evolve
with time in such a cosmological scenario. Changes in the hole
size are expected to be very much magnified in case that dark
energy be phantom energy and will be also considered in the
present paper. In addition, we shall explore as well the
interesting possibility that the Lorentzian holes might be used in
the future to circumvent the big rip. The main results of this
endeavor are that, in fact, relative to an observer in the initial
embedding-space coordinate system, whereas the shape of the holes
remains constant along current expansion, their size increases
with time at a rate proportional to the cosmic scale factor, and
that one might envisage a kind of scenarios where sufficiently
enlarged holes could be used by future advanced civilizations to
escape towards their future from the big rip.

The paper is outlined as follows. In Sec. II we solve the
Friedmann equations for an accelerating universe endowed with
quintessential dark energy, with and without a positive
cosmological constant, where the regime of phantom energy is also
considered. The effects of accelerated and "super-accelerated"
expansion of the universe on the size and shape of wormholes and
ringholes are investigated in detail in Sec. III. Sec. IV contains
a calculation of the stress-energy tensor for both wormholes and
ringholes. For the former hole the "exoticity" function [23] is
explicitly derived and analyzed. A rather speculative scenario is
considered in Sec. V in which wormholes and ringholes are used by
future advanced civilizations in order to circumvent the
singularity at the big rip. We finally summarize and conclude in
Sec. VI, where some short comments are added.

\section{Cosmological solutions for a dark-energy vacuum}

In this section I shall derive solutions to the Friedmann
equations that correspond to a quintessence scalar field $\phi$,
assuming an equation of state with perfect-fluid form
\begin{equation}
p=\omega\rho,
\end{equation}
where, consistent with present constraints [16], the parameter
$\omega$ is assumed to be constant and to take on values within
the interval $-2/3>\omega\geq -3/2$, and the pressure $p$ and the
energy density $\rho$ are defined by [17]
\begin{equation}
p=\frac{1}{2}\dot{\phi}^2-V(\phi)
\end{equation}
\begin{equation}
\rho=\frac{1}{2}\dot{\phi}^2+V(\phi),
\end{equation}
with $V(\phi)$ the potential energy for the quintessence field
$\phi$ whose precise form is not required in what follows. We see
that for sufficiently slowly-varying field $\phi$, we have that
$\dot{\phi}^2 < V(\phi)$, so that the equation-of-state parameter
can take on the values specified by the above $\omega$-interval.
Inserting then the relation $p=\omega\rho$ in the general
expression for cosmic energy conservation, $d\rho=-3(p+\rho)da/a$
(in which $a$ is the scale factor) and integrating, we obtain for
the energy density
\begin{equation}
\rho=Ra^{-3(1+\omega)},
\end{equation}
where $R$ is an integration constant which should be real and
positive. We shall consider in what follows two cases. We regard
the cosmic vacuum dark energy to either solely contain the
quintessence field $\phi$, or be made by the combination of the
quintessence field $\phi$ plus a pure positive cosmological
constant $\Lambda$. In the first case, the Friedmann equation
reads
\begin{equation}
\left(\frac{\dot{a}}{a}\right)^2=Aa^{-3(1+\omega)},
\end{equation}
with $A=8\pi GR/3$. This differential equation can be easily
integrated to yield:
\begin{equation}
a(t)=\left[a_0^{3(1+\omega)/2}+
\frac{3(1+\omega)\sqrt{A}}{2}(t-t_0)\right]^{2/[3(1+\omega)]} ,
\end{equation}
where $a_0$ and $t_0$ are the initial radius and time,
respectively. Note that for $\omega >-1$ this solution describes
an accelerating universe whose scale factor increases towards
infinity as $t\rightarrow\infty$. The case for $\omega <-1$ is
even more interesting. It corresponds to the so-called phantom
dark energy for which the dominant energy condition is violated,
i.e. [22]
\begin{equation}
p+\rho <0 ,
\end{equation}
even though the energy density is surprisingly ever increasing.
Notice furthermore that in this case the scale factor blows up at
a finite time,
\begin{equation}
t_* =t_0+\frac{2}{3(|\omega|-1)a_0^{3(|\omega|-1)/2}} ,
\end{equation}
giving rise to what is now known as a "big rip" [22].

In the most general case where we add to the cosmic quintessence
vacuum a positive cosmological constant $\Lambda$, the Friedmann
equation would be
\begin{equation}
\left(\frac{\dot{a}}{a}\right)^2=\lambda+Aa^{-3(1+\omega)},
\end{equation}
where $\lambda=\Lambda/3$. This equation admits also an analytical
solution in closed form [24]
\begin{eqnarray}
&&a(t)=\left(\frac{A}{4C\lambda}\right)^{1/[3(1+ \omega)]}\times
\left(e^{\frac{3(1+\omega)\sqrt{\lambda}(t-t_0)}{2}}\right.\nonumber\\
&&\left.-Ce^{\frac{-3(1+
\omega)\sqrt{\lambda}(t-t_0)}{2}}\right)^{2/[3(1+\omega)]} ,
\end{eqnarray}
where
\begin{equation}
C=\frac{\sqrt{\lambda+Aa_0^{-3(1+\omega)}}-
\sqrt{\lambda}}{\sqrt{\lambda+Aa_0^{-3(1+\omega)}}
+\sqrt{\lambda}},
\end{equation}
with $a_0$ the initial scale factor. Since $0<C<1$ we have also in
this case the possibility of a big rip at time
\begin{equation}
t_*=t_0 -\frac{\ln C}{3(|\omega|-1)\sqrt{\lambda}}
\end{equation}
in the regime where $\omega<-1$. One would expect that in the
vecinity of time $t_*$ the size of wormholes and ringholes
increased in a rather dramatic way in the two considered cases. We
see that in both of such cases the larger $a_0$ and $|\omega|$ the
nearer the doomsday, and hence the time expected to have
macroscopic wormholes and ringholes in our universe.

\section{Growth of holes in an accelerating universe}

In this section we shall investigate the way in which the size and
form of Lorentzian tunnelings, which can be given as topological
four-dimensional generalizations of the Misner space [6] embedded
in the current universe, can be affected by the present
accelerating expansion and, in particular, the possibility that
there will be a big rip. We shall restrict ourselves to consider
two cases. On the one hand, the analysis done by Roman [14] for
inflating Morris-Thorne wormhole [3] will here be adapted to study
the case of accelerating wormholes. On the other hand, we will
also investigate the subtleties that would arise when considering
accelerating ringholes, where the tunneling mouths have the
topology of a torus [4].

\subsection{Morris-Thorne wormhole}

The Lorentzian metric of the Morris-Thorne wormhole can be given
as [3]
\begin{equation}
ds^2= -e^{2\Phi(r)}dt^2 +\frac{dr^2}{1-\frac{K(r)}{r}}+r^2
d\Omega_2^2 ,
\end{equation}
where $\Phi$ may be taken to be either zero or a given function of
the radial coordinate $r$ otherwise, the function $K(r)$ can be
taken either as $K(r)=K_0^2/r$ for wormhole with zero tidal force,
or as $K(r)=K_0[1-(1-K_0)/R_0]^2$ if the exotic matter is confined
into an arbitrarily small region around the wormhole throat, and
$d\Omega_2^2= d\theta^2+\sin^2\theta d\phi^2$ is the metric on the
unit two-sphere.

Following Roman [14], we shall now generalize the static metric
(3.1) to a time-dependent background metric that describes the
time-evolution of an initially static wormhole with metric (3.1)
immersed in an expanding universe filled with a quintessence field
with or without a cosmological constant. This will be done by
simply inserting a dimensionless factor proportional to the square
of the scale factor into the three-dimensional spatial part of
metric (3.1). Taking for that factor
\begin{equation}
g(t)^2=\left(1+\frac{3(1+\omega)\sqrt{A}(t-
t_0)}{2a_0^{3(1+\omega)/2}}\right)^{2/[3(1+\omega)]}
\end{equation}
for the pure quintessential case, and
\begin{equation}
g(t)^2=\left(e^{\frac{3(1+\omega)\sqrt{\lambda}(t-t_0)}{2}}
-Ce^{\frac{-3(1+
\omega)\sqrt{\lambda}(t-t_0)}{2}}\right)^{2/[3(1+\omega)]},
\end{equation}
for the case in which we add a positive cosmological constant to
the quintessence field, we have for the time-dependent metric of a
wormhole in an accelerating universe:
\begin{equation}
ds^2= -e^{2\Phi(r)}dt^2
+g(t)^2\left(\frac{dr^2}{1-\frac{K(r)}{r}}+r^2 d\Omega_2^2\right)
.
\end{equation}
Here the coordinates on the two-sphere have the same geometrical
interpretation as for the Morris-Thorne case, and circles of
constant $r$ are centered on the throat of the wormhole. Following
again Roman [14], the coordinate system is chosen to be comoving
with the wormhole geometry, with the throat always located at
$r=K_0$. Of course, in the limit where $K(r)=\Phi(r)=0$, metric
(3.4) reduces to the FRW metric for an accelerating universe
driven by a quintessence field with or without a cosmological
constant (see Sec. II), and for $A=0$ and $\lambda=0$ it reduces
to the wormhole metric (3.1). If we just set $A=0$ then we would
recover the case of an inflating wormhole considered by Roman. We
note furthermore that the spacetime described by metric (3.4) is
inhomogeneous because of the presence of the wormhole.

Let us next consider how an originally nearly Planck-sized
wormhole is being enlarged by the accelerating expansion of the
universe. The initial wormhole would be a reasonable wormhole at
the onset of dark energy dominance at $t=t_0$ for a suitable
choice of functions $K(r)$ and $\Phi(r)$, which we shall assume.
Considering then the proper circumference $c$ at the wormhole
throat, $r=K=K_0$, for $\theta=\pi/2$ and any constant time, we
have in the accelerating framework
\begin{equation}
c=K_0\int_0^{2\pi}d\phi g(t)=2\pi K_0 g(t) ,
\end{equation}
which simply is $g(t)$ times the initial circumference, with
$g(t)$ given by Eq. (3.2) or Eq. (3.3), depending on whether we do
not add or we do add a cosmological constant to the quintessence
field. A similar conclusion is also obtained for radial proper
length through the wormhole between any two points $A$ and $B$ at
$t=$const., as that quantity can be computed to be
\begin{equation}
d(t)=\pm g(t)\int_{r_A}^{r_B}\frac{dr}{\sqrt{1-K(r)/r}},
\end{equation}
which for the simplest wormhole where $K(r)=K_0^2/r$ becomes,
\[d(t)=\pm g(t)\left(\sqrt{r_B^2-K_0^2}-
\sqrt{r_B^2-K_0^2}\right) .\] Thus, both the size of the throat
and the radial proper distance between the wormhole mouths
increase with $t$ at exactly the same accelerating rate as that of
the universe expansion.

That the form of the wormhole metric is preserved with
cosmological time can also be easily seen by e.g. using the
embedding [14] of a $t=$const., $\theta=\pi/2$ slice of the
spacetime (3.4) in a flat three-dimensional Euclidean space with
cylindrical metric
\begin{equation}
ds^2=d\bar{z}^2+d\bar{r}^2+\bar{r}^2 d\phi^2 .
\end{equation}
Since the metric on the chosen slice is
\begin{equation}
ds^2= g(t)^2\left(\frac{dr^2}{1-K(r)/r}+r^2 d\phi^2\right) ,
\end{equation}
we get
\begin{equation}
\bar{r}=g(t)r|_{t={\rm const.}}
\end{equation}
\begin{equation}
d\bar{r}^2=g(t)^2 dr^2|_{t={\rm const.}}.
\end{equation}
It follows that, relative to the $\bar{z},\bar{r},\phi$
coordinates, the form of the wormhole metric will be preserved
provided that the metric on the embedded slice has the form
\begin{equation}
ds^2= \frac{d\bar{r}^2}{1-\bar{K}(\bar{r})/\bar{r}}+\bar{r}^2
d\phi^2 ,
\end{equation}
in which $\bar{K}(\bar{r})$ must have a minimum at some
$\bar{K}_0=\bar{r}_0$, such as it occurred in the inflating
wormhole case [14]. By using Eqs. (3.9) and (3.10) it is easy to
see that Eqs. (3.8) and (3.11) can be re-written into each other
if we take
\begin{equation}
\bar{K}(\bar{r})=g(t)K(r) ,
\end{equation}
which is satisfied by any of the choices for $K(r)$ made explicit
by Morris and Thorne [3]. It follows that also a wormhole evolving
in a universe filled with dark energy will have the same overall
size and shape relative to the $\bar{z},\bar{r},\phi$ coordinate
system as the initial wormhole had relative to the initial
$z,r,\phi$ embedding-space coordinate system. This can be seen if
we consider the embedding procedure used by Morris and Thorne;
i.e. if we take as the metric of the embedding space
\begin{equation}
ds^2=dz^2+dr^2+r^2 d\phi^2 .
\end{equation}
Now, since the embedded surface is axially symmetric, $z=z(r)$ and
the metric on the embedded surface can be expressed as
\begin{equation}
ds^2=\left[1+\left(\frac{dz}{dr}\right)^2\right]dr^2 +r^2 d\phi^2
.
\end{equation}
Metrics (3.13) and (3.14) can be identified if (i) the coordinates
$r,\phi$ of the embedding space and of the wormhole are
identified, and (ii) we require
\begin{equation}
\frac{dz}{dr}=\pm\left(\frac{r}{K(r)}-1\right)^{-1/2} .
\end{equation}
Applying this procedure to our present case, we have
\begin{equation}
\frac{d\bar{z}}{d\bar{r}}=
\pm\left(\frac{\bar{r}}{\bar{K}(\bar{r})}-1\right)^{-1/2}=
\frac{dz}{dr} ,
\end{equation}
and hence,
\begin{eqnarray}
&&\bar{z}(\bar{r})
=\pm\int\frac{d\bar{r}}{\sqrt{\frac{\bar{r}}{\bar{K}(\bar{r})}
-1}}\nonumber\\ &&\pm g(t)\int\frac{dr}{\sqrt{\frac{r}{K(r)}- 1}}
=g(t)z(r) ,
\end{eqnarray}
again as in the Roman's inflating wormhole [14]. Using finally
Eqs. (3.10) and (3.17) we then also obtain the same relation
between our embedding space at any time $t$ and the initial
embedding space at $t=0$; i.e.
\begin{equation}
ds^2= d\bar{z}^2+d\bar{r}^2+\bar{r}^2 d\phi^2
=g(t)^2\left(dz^2+dr^2+r^2 d\phi^2\right) .
\end{equation}
Using Eqs. (3.9), (3.10), (3.12) and (3.16) we also obtain for the
Morris-Thorne flareout condition [3],
\begin{eqnarray}
&&\frac{d^2\bar{r}(\bar{z})}{d\bar{z}^2}
=g(t)^{-1}\left(\frac{K-K'r}{2K^2}\right)\nonumber\\
&&g(t)^{-1}\left(\frac{d^2 r(z)}{dz^2}\right)=
\frac{\bar{K}-\bar{K}'\bar{r}}{2\bar{K}^2} > 0,
\end{eqnarray}
at or near the throat. In obtaining this condition use has been
made of $\bar{K}(\bar{r})'=d\bar{K}/d\bar{r}=K(r)'=dK/dr$.

The conclusion is thus obtained that, whereas the shape of a
wormhole naturally existing in an accelerating universe is
preserved with time, its size is increased relative to the initial
embedding-space coordinate system. We note that if the equation of
state for dark energy is characterized by a parameter $\omega>-1$,
then the size increase will be slower than in inflating universe,
but if the wormhole is placed in a universe filled with phantom
energy for which $\omega <-1$ (which is the most natural situation
for a wormhole to exist because in this case the required energy
exoticity for the wormhole existence is shared by the overall
universe), then the wormhole will increase size more rapidly than
in the inflating case, which is characterized by the parameter
$\omega=-1$ that corresponds to a cosmological constant.

\subsection{Ringhole}

The case of a ringhole entails certain subtleties that make the
treatment of its time-dependence in an accelerating universe
somehow different of the case of an inflating wormhole. A ringhole
is nothing but a wormhole with the spherical topology being
replaced for a toroidal topology, that is it is just a
four-dimensional Misner space where the planes are substituted by
tori whose relevant geometric parameters are given in Fig. 1 (a).
The embedding of a ringhole in flat regions of the universe is
depicted in Fig. 1 (b), where we point out the throat position and
how the inner and outer surfaces flare respectively inward and
outward near the throat. The general Lorentzian metric for a
ringhole can be written as
\begin{eqnarray}
&&ds^2= -e^{2\Phi(r)}dt^2 +\frac{dr^2}{1-\frac{K(b)}{b}}+m^2
d\phi_1^2+b^2 d\phi_2^2\nonumber\\ &&=-e^{2\Phi(r)}dt^2
+\frac{n^2}{r^2}d\ell^2+m^2 d\phi_1^2+b^2 d\phi_2^2 ,
\end{eqnarray}
where $\phi_1$ and $\phi_2$ are the angles which usually define
the position on the torus (See Fig. 1 (a)),
\begin{equation}
b=\sqrt{b_0^2+\ell^2}
\end{equation}
\begin{equation}
m=a-b\cos\phi_2
\end{equation}
\begin{equation}
n=b-a\cos\phi_2
\end{equation}
\begin{equation}
r=\sqrt{a^2+b^2-2ab\cos\phi_2} ,
\end{equation}
$a$ is the constant radius of the axial circumference defined by
the torus, and $b$ is the radius of the circular transversal
section on the torus, with $b_0$ giving that radius at the throat.

Following the same procedure as for the Morris-Thorne wormhole, we
can now write the ringhole metric generalized to a time-dependent
accelerating dark-energy background in the form
\begin{equation}
ds^2= -e^{2\Phi(r)}dt^2
+g(t)^2\left(\frac{dr^2}{1-\frac{K(b)}{b}}+m^2 d\phi_1^2+b^2
d\phi_2^2\right) ,
\end{equation}
with $g(t)$ as given by either Eq. (3.3) or Eq. (3.2 ), depending
on whether we introduce or not a cosmological constant among the
components of dark energy. Because of the presence of the
ringhole, the spacetime described by metric (3.25) is
inhomogeneous as well. Also in this case the metric reduces to the
FRW metric for an accelerating universe in the limit
$K(b)=\Phi(r)=0$, and becomes the ringhole metric (3.20) for
$A=0$, $\lambda=0$ simultaneously. If we would just set $A=0$,
then we have an inflating ringhole evolving in a de Sitter-like
space.

In what follows we apply the analysis done for wormholes to the
case of ringholes. In the latter case we have two proper
circumferences, $c_{{\rm min}}$ and $c_{{\rm max}}$ at the throat.
They are defined by: $c_{{\rm min}}: \ell=0, \phi_2=0$, and
$c_{{\rm max}}: \ell=0, \phi_2=\pi$, both at any time $t=$const.
We then have
\begin{equation}
c_{{\rm min}}=\int_0^{2\pi}g(t)(a-b_0)d\phi_1=2\pi g(t)r_{{\rm
min}}=2\pi g(t)m_{{\rm min}}
\end{equation}
\begin{equation}
c_{{\rm max}}=\int_0^{2\pi}g(t)(a+b_0)d\phi_1=2\pi g(t)r_{{\rm
max}}=2\pi g(t)m_{{\rm max}} ,
\end{equation}
where the subscript "min" denotes $\phi_2=0$ and the subscript
"max" denotes $\phi_2=\pi$, in the corresponding expressions for
$m$ and $r$. Thus, one obtains that both circumferences are
equally increased by a factor $g(t)$.

The radial proper length through a ringhole between any two points
$A$ and $B$ at any constant time and constant $\phi_2$ is
similarly increased by a factor $g(t)$ generally according to:
\begin{equation}
d(t)=\pm g(t)\int_{r_A}^{r_B}\frac{dr}{\sqrt{1- \frac{K(b)}{b}}}
\end{equation}
which once again is just $g(t)$ times the initial radial proper
separation. In case of separation along the ringhole inner surface
$\phi_2=0$, if we assume the simplest ringhole choice $K=K_0^2/b$
such that $K_0 <<(a-r)$, we obtain the approximate expression
\begin{equation}
d(t)\simeq g(t)\left(r_B -r_A\right)\left[1+
\frac{K_0^2}{2\left(a-r_B\right)\left(a-r_A\right)}\right] .
\end{equation}
For the same ringhole choice but now with $K_0 <<(r-a)$, we would
have for separation distances along the outer surface $\phi=\pi$,
\begin{equation}
d(t)\simeq g(t)\left(r_B -r_A\right)\left[1+
\frac{K_0^2}{2\left(r_B-a\right)\left(r_A-a\right)}\right] .
\end{equation}
The conclusion is then: Both, the distinct geometrical sizes
defined on the rinhole throat and the radial proper distances
defined along any rinhole surface between the two mouths increase
with time following the same pattern.

In order to investigate whether the form of a ringhole is also
preserved with time in the accelerating background, let us again
embed in a flat three-dimensional Euclidean space with metric
(3.13) either a slice $t=$const., $\phi_2=0$, or a slice
$t=$const., $\phi_2=\pi$. On the first of such slices one would
have
\begin{equation}
ds^2= g(t)^2 \left(\frac{dr^2}{1-\frac{K(b)}{b}}+m^2
d\phi_1^2\right).
\end{equation}
Comparing Eq. (3.13) with Eq. (3.31), we have
\begin{equation}
\bar{r}=\bar{m}=g(t)r|_{t={\rm const.}}=g(t)m|_{t={\rm const.}},
\end{equation}
\begin{equation}
d\bar{r}=d\bar{m}=g(t)dr|_{t={\rm const.}}= g(t)dm|_{t={\rm
const.}} ,
\end{equation}
with
\[r=r_{{\rm min}}=m=m_{{\rm min}}=a-b . \]
On the other extreme slice $\phi_2=\pi$ we obtain the same
expressions (3.32) and (3.33), but now referred to the maximal
values of parameters $r$ and $m$, i.e.
\[r=r_{{\rm max}}=m=m_{{\rm max}}=a+b . \]
It follows that expressions (3.32) and (3.33) are valid for any
slice $t=$const., from $\phi_2=0$ to $\phi_2=\pi$. Hence,
\begin{equation}
\bar{n}=g(t)n ,\;\;\; \bar{a}=g(t)a ,\;\;\; \bar{b}=g(t)b ,
\end{equation}
for any of such slices.

Thus, with respect to the $\bar{z},\bar{r},\phi$ coordinates, the
form of the ringhole metric will be preserved if the metric on the
embedded slice is written as
\begin{equation}
ds^2= \frac{d\bar{r}^2}{1 -\frac{\bar{K}(\bar{b})}{\bar{b}}}
+\bar{m}^2 d\phi_1^2 ,
\end{equation}
where $\bar{K}(\bar{b})$ is assumed to have a minimum at some
$\bar{K}(\bar{b}_0)=\bar{K}_0$. We can re-write Eq. (3.35) in the
form
\begin{equation}
ds^2= g(t)^2 \left(\frac{dr^2}{1 -\frac{K(b)}{b}} +m^2
d\phi_1^2\right) ,
\end{equation}
by using Eqs. (3.32) - (3.34) and
\begin{equation}
\bar{K}(\bar{b})=g(t)K(b) .
\end{equation}
Choosing for the ringhole one which is defined to contain exotic
material everywhere, that is
\begin{equation}
K(b)=K_0^2/b ,
\end{equation}
and re-writing the ring-hand-side of expression (3.38) in terms of
$\bar{b}$ as given in Eq. (3.34), by using Eq. (3.32) we can see
that expression (3.37) is satisfied provided the third of Eqs.
(3.34) is satisfied, and {\it vice versa}.

It follows that the enlarged ringhole will also have the same
overall size and shape relative to coordinates
$\bar{z},\bar{r},\phi$, as the initial ringhole had relative to
the initial $z,r,\phi$ embedding space coordinates. This can be
most clearly seen by pointing out that our embedding scheme
corresponds to one whose $z,r$ coordinates scale with time, such
as it can be shown following the embedding procedure given in Ref.
[4] for Eqs. (3.7) and (3.35). Using the above expressions for
relating the ringhole parameters, one can get
\begin{equation}
\frac{d\bar{z}}{d\bar{r}}=\pm
\left(\frac{\bar{b}}{\bar{K}(\bar{b})}-1\right)^{-1/2}=\frac{dz}{dr}
,
\end{equation}
which implies
\begin{eqnarray}
&&\bar{z}(\bar{r})=\pm
\int\frac{d\bar{r}}{\sqrt{\frac{\bar{b}}{\bar{K}(\bar{b})}
-1}}\nonumber\\ &&=\pm g(t)\int\frac{dr}{\sqrt{\frac{b}{K(b)} -1}}
=g(t)z(r) .
\end{eqnarray}
Note that if we take for the ringhole a homogeneous distribution
of exotic matter, then on the surface $\phi_2=0$,
\[z(r)=\mp K_0\ln\left(b+\sqrt{b^2-K_0^2}\right) ,\]
and the same expression with reversed sign on the surface
$\phi_2=\pi$. Therefore the relation between our embedding space,
at any time $t$ and any value of the angular coordinate $\phi_2$,
and the initial embedding space at $t=0$ and the same chosen value
of $\phi_2$ has the same form as that is given by Eq. (3.18) for
wormholes. Thus, relative to the barred coordinates, the ringhole
does not change size because the scaling of the embedding space
exactly compensates the expansion of the ringhole, such as it also
occurred for wormholes.

The analogous of the flareout and flarein conditions of the
initial ringhole [4], are respectively given at any nonzero time
$t$ by
\begin{equation}
\frac{d^2\bar{r}}{d\bar{z}^2}>0 ,\;\;\; 2\pi-\phi_2^c >\phi_2
<\phi_2^c
\end{equation}
\begin{equation}
\frac{d^2\bar{r}}{d\bar{z}^2}<0 ,\;\;\; -\phi_2^c >\phi_2
<\phi_2^c ,
\end{equation}
at or near the throat, with $\phi_2^c = \arccos(b/a)$ at the
angular horizons [4]. From the set of the above equations, one has
\begin{eqnarray}
&&\frac{d^2\bar{r}}{d\bar{z}^2}
=\frac{\bar{r}\bar{K}(\bar{b})-\bar{K}(\bar{b})'
\bar{b}\bar{n}}{\bar{n}\bar{K}(\bar{b})^2}\nonumber\\&&=\frac{d^2
r(z)}{g(t)dz^2}= \frac{rK(b)-K(b)' bn}{g(t)nk(b)^2} >0 ,
\end{eqnarray}
for the flareout condition, and
\begin{eqnarray}
&&\frac{d^2\bar{r}}{d\bar{z}^2}
=\frac{\bar{r}\bar{K}(\bar{b})-\bar{K}(\bar{b})'
\bar{b}\bar{n}}{\bar{n}\bar{K}(\bar{b})^2}\nonumber\\&&=\frac{d^2
r(z)}{g(t)dz^2}= \frac{rK(b)-K(b)' bn}{g(t)nk(b)^2} <0 ,
\end{eqnarray}
for the flarein condition. In these expressions
$\bar{K}'=d\bar{K}/d\bar{r}$ and $K'=dK/dr$. It follows that the
flareout and flarein conditions relative to the barred quantities
have the same form as those for static ringholes. With respect to
the non-barred quantities, whereas the flareout condition appears
as though it might be harder to satisfy as time goes on, the
flarein condition looks as though it were easier to satisfy as
time goes on. This is however due to the fact that as the ringhole
increases, both its throat size and any proper length also
increase along with the surrounding space. The ringhole then needs
to flareout less and less on the outer surfaces near the throat,
and to flarein more and more on the inner surfaces near the
throat, as the two external spaces which are connected by the
ringhole move farther apart.

We thus again obtain for ringholes in an accelerating universe the
conclusion that, relative to the initial embedding-space
coordinate system, the size of the hole will increase with time
without undergoing any deformation whatsoever. If dark energy is
governed by an equation of state with parameter $\omega>-1$, then
that increase of the hole size will be once again slower than for
inflating holes, but if $\omega<-1$ the ringhole will
"super-inflate", with the smaller $\omega$, the larger the
expansion rate of the ringhole. Since it was showed [4] that, even
near their throat, the internal channel of these ringholes possess
certain itineraries through which a traveler would not find any
exotic matter, one could envisage a big-rip scenario for which
initially submicroscopic ringholes could be grown to macroscopic
sizes through which future inhabitants of the universe might find
themselves saved just before the big rip started to smash their
worlds. A more detailed proposal about this possibility will be
considered in Sec. V.

\begin{figure}
\includegraphics[width=.9\columnwidth]{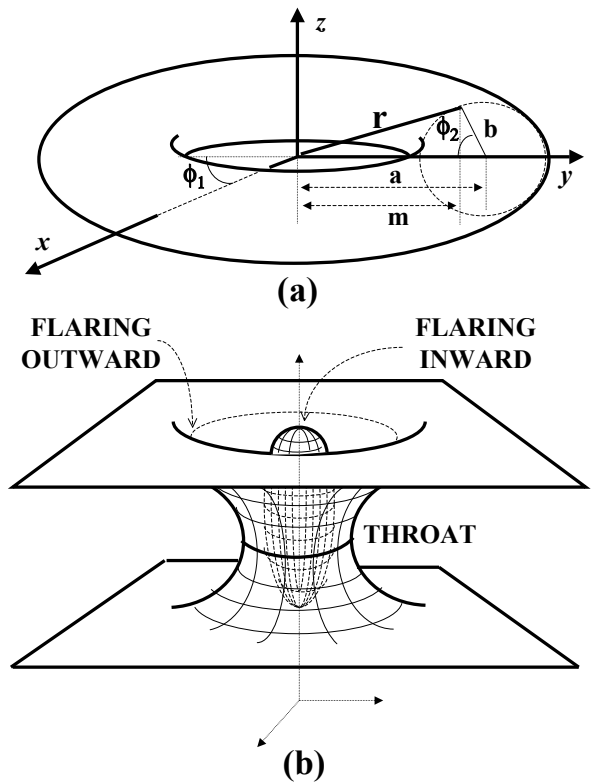}
\caption{\label{fig:epsart} (a) Coordinates and geometric
parameters on the two-torus; any point on the torus surface can be
labeled by parameters $a$, $b$, $\phi_1$ and $\phi_2$. (b)
Embedding diagram for the spacetime of a ringhole connecting two
asymptotically flat regions.}
\end{figure}

\section{The stress-energy tensor and exoticity}

Using the notation that stems from defining the set of orthonormal
basis vectors,
\begin{equation}
e_{\hat{t}}=e^{-\Phi}e_t ,\;\;
e_{\hat{r}}=\frac{\sqrt{1-K(r)/r}}{g(t)}e_r \nonumber
\end{equation}
\begin{equation}
e_{\hat{\theta}}=\frac{e_{\theta}}{rg(t)},\;\;
e_{\hat{\phi}}=\frac{e_{\phi}}{g(t)r\sin\theta} ,
\end{equation}
for wormholes, and
\begin{equation}
e_{\hat{t}}=e^{-\Phi}e_t ,\;\;
e_{\hat{r}}=\frac{\sqrt{1-K(b)/b}}{g(t)}e_r \nonumber
\end{equation}
\begin{equation}
e_{(\hat{\phi})_1}=\frac{(e_{\phi})_1}{mg(t)},\;\;
e_{(\hat{\phi})_2}=\frac{(e_{\phi})_2}{bg(t)} ,
\end{equation}
for ringholes, from the Einstein equation,
\[R_{\hat{\mu}\hat{\nu}}-\frac{1}{2}g_{\hat{\mu}\hat{\nu}}R =8\pi
T_{\hat{\mu}\hat{\nu}} ,\] we obtain the components of the stress
energy tensor (incorporating in them the-dark energy vacuum
contributions). In the most general case that we take for $g(t)$
expression (3.3) for an expanding wormhole, these components are
\begin{equation}
T_{\hat{t}\hat{t}}= \frac{1}{8\pi [-]^2}\left(\frac{4}{3}C_1^2
[+]^2 +\frac{K'}{r^2}\right)=\rho(r,t)
\end{equation}
\begin{eqnarray}
&&T_{\hat{r}\hat{r}} =\frac{1}{8\pi}\left\{-\frac{2}{3}C_1^2
\left(\frac{\frac{5}{3}[+]^2}{[-]^2}
+(1+\omega)\right)\right.\nonumber\\
&&\left.-\frac{\frac{K}{r^2}-2\Phi'(1
-\frac{K}{r})}{r[-]^{\frac{4}{3(1+\omega)}}}\right\} =-\tau(r,t)
\end{eqnarray}
\begin{equation}
T_{\hat{t}\hat{r}} =\frac{4C_1[+]e^{-\Phi}\Phi'\sqrt{1-
\frac{K}{r}}}{24\pi [-]^{\frac{5+3\omega}{3(1+\omega)}}} =-f(r,t)
\end{equation}
\[8\pi T_{\hat{\theta}\hat{\theta}}=8\pi T_{\hat{\phi}\hat{\phi}}=8\pi p(r,t)=\]
\begin{eqnarray}
&&\left\{\frac{4}{3}C_1^2\left(\frac{\omega[+]^2}{[-]^2}
-(1+\omega)\right)e^{-2\Phi}\right.\nonumber\\
&&\left.-[-]^{-\frac{4}{3(1+ \omega)}}
\left[\frac{1}{2}\left(\frac{K}{r^3}- \frac{K'}{r^2}\right)
+\frac{\Phi'}{r}\left(1-\frac{K}{2r}
-\frac{K'}{2}\right)\right.\right.\nonumber
\end{eqnarray}
\begin{equation}
\left.\left.+\left(1-\frac{K}{r}\right)\left(\Phi''+
(\Phi')^{2}\right)\right]\right\}=p(r,t),
\end{equation}
where
\begin{equation}
[\pm]=e^{C_1(1+\omega)t}\pm C_2 e^{C_1(1+\omega)t} .
\end{equation}

In the case of a ringhole and $g(t)$ given by Eq. (3.3), we get
instead
\begin{equation}
T_{\hat{t}\hat{t}}=T_{\hat{t}\hat{t}}(t=0) + \frac{C_1^2
[+]^2}{12\pi [-]^2} = \rho(r,t)
\end{equation}
\begin{equation}
T_{\hat{r}\hat{r}}=T_{\hat{r}\hat{r}}(t=0)
-\frac{C_1^2}{12\pi}\left(\frac{\frac{5}{3}[+]^2}{[-]^2}
+(1+\omega)\right)=-\tau(r,t)
\end{equation}
\begin{equation}
T_{\hat{t}\hat{r}}=T_{\hat{t}\hat{r}}(t=0)+ \frac{C_1
e^{-\Phi}\Phi' \sqrt{1-\frac{K}{b}}[+]}{6\pi
[-]^{\frac{5+3\omega}{3(1+\omega)}}}=-f(r,t)
\end{equation}
\begin{eqnarray}
&&T_{\hat{\phi_1}\hat{\phi_1}}= T_{\hat{\phi_2}\hat{\phi_2}}=
T_{\hat{\phi_1}\hat{\phi_1}}(t=0)\nonumber\\ &&+ \frac{C_1^2
e^{-2\Phi}}{6\pi}\left(\frac{\omega[+]^2}{[-]^2}
-(1+\omega)\right)=p(r,t) ,
\end{eqnarray}
where [+] and [-] are time-dependent functions which are again
defined in Eq. (4.9). In these equations $\rho, \tau, f$ and $p$
are the mass-energy density, the radial tension per unit area, the
energy flux in the radial direction and the lateral pressures, as
measured by observers stationed at constant values of the
coordinates ($r,\theta,\phi$ or $r,\phi_1,\phi_2$), respectively.
The terms denoted by $T_{\hat{\mu}\hat{\nu}}(t=0)$ for the
ringhole case correspond to those of the static case and can
easily be derived from those provided in Ref. [4]. The energy
fluxes in the radial direction given by Eqs. (4.7) and (4.12) for
wormholes and ringholes, respectively, will vanish at the throat
in both cases, such as one should expect. More surprising is the
feature that all the components of the stress-energy tensor will
diverge as one approaches the critical big-rip time, $t\rightarrow
t_*$, if $\omega<-1$, both for wormholes and ringholes. This is an
interesting result that is related with the known fact that vacuum
energy density increases with time, tending to infinity at the big
rip, for phantom-energy quintessential models [20]. Assuming, on
the other hand, that $\Phi(r)\rightarrow 0$ and either
$K/r\rightarrow 0$ or $K/b\rightarrow 0$, as $r\rightarrow\infty$,
the stress-energy components asymptotically recover their pure
cosmological dark-energy forms which also contained a cosmological
constant. Had we used the expansion factor given by Eq. (3.2) in
assuming that limit, then we had obtained stress-energy components
in terms of the quintessence-field parameters only. Finally, for
the particularly simple example that one chooses holes such that
$\Phi(r)=0$ everywhere, we see that the energy flux in the radial
direction becomes identically zero for both kinds of holes, as it
should also be expected.

In what follows of the present section we shall restrict ourselves
to briefly discuss the exotic character of the matter entering the
holes. For the sake of simplicity, we confine our analysis to the
case of wormholes, but note that all the results we obtain for
this case are directly applicable to the ringhole case. In order
to describe the exotic nature of the matter in these holes, an
"exoticity" function has been introduced which is generally
defined by [23]
\begin{equation}
\xi=\frac{\tau-\rho\mp f}{|\rho|} .
\end{equation}
In the case of a wormhole the exoticity function can be written in
the explicit form:
\begin{eqnarray}
&&\xi= \left\{-\left[-\frac{2}{3}C_1^2\left(\frac{5}{3}[+]^2
+(1+\omega)[-]^2\right) -\Psi_1(r)\right]\right.\nonumber\\
&&\left. -\left(\frac{4}{3}C_1^2[+]^2+
\frac{K'}{r^2}\right)\right\}\left|\frac{4}{3}C_1^2[+]^2
+\frac{K'}{r^2}\right|^{-1}\pm \Psi_2(r) ,\nonumber
\end{eqnarray}
\begin{equation}
\end{equation}
where
\begin{equation}
\Psi_1=\left(\frac{K}{r^2}-22\Phi'(1-\frac{K}{r})[-]^{\frac{2(1+
3\omega)}{3(1+\omega)}}\right)
\end{equation}
\begin{equation}
\Psi_2=\frac{4e^{-\Phi}\Phi'\sqrt{1 -\frac{K}{r}}C_1
[+][-]^{\frac{1+3\omega}{3(1+\omega)}}}{3\left|\frac{4}{3}C_1^2
[+]^2+\frac{K'}{r^2}\right|} .
\end{equation}
Eq. (4.15) is generally nonzero, and hence the exotic character of
the matter in the wormhole is also generally preserved along
evolution. In case that dark energy is phantom energy, i.e. if
$\omega<-1$, however, as one approaches the big rip at
$t=t_*=-\ln(C_2)/[2C_1 (|\omega|-1)]$, Eq. (4.15) reduces to the
approximate expression
\begin{equation}
\xi_{{\rm ph}}\simeq\frac{\frac{K'}{r^2}-\frac{16}{9}C_1^2
C_2}{\left|\frac{K'}{r^2}+\frac{16}{3}C_1^2 C_2\right|} .
\end{equation}
We can now see that the matter entering a wormhole approaching the
big rip may be of two different kinds, as related to the
cosmological parameters of the universe where the hole is embedded
in. If the wormhole-defining parameter $K\neq K_0+16C_1^2 C_2r^3$,
where $K_0$ is an arbitrary constant, then the matter is still
exotic in the neighborhood of the big rip. However, if the
wormhole is defined so that $K\simeq K_0+16C_1^2 C_2r^3$, then
there will be a time as we approach the big rip at which all the
matter in the wormhole behaved like though it were ordinary
matter, without any violation of the energy conditions. As it was
pointed out before, these results also apply to the case of a
ringhole when it approaches the big rip.

\section{Escaping towards the future from big rip?}

The content of this section has a rather speculative character. We
shall deal with the big rip scenario, suggested by Caldwell,
Kamionkowsni and Weinberg [22], for any of the two cosmological
scenarios considered in Sec. II, whenever the parameter of the
equation of state is $\omega<-1$. Depending on the precise values
assumed for $\omega$ and the initial size $a_0$, as the doomsday
is being approached, one would expect astrophysical objects, such
as clusters and galaxies, next the solar system and then the Earth
itself and all its constituents and living inhabitants, to
successively disappear beyond the horizon of any observer. At the
big rip not even quarks and electrons are left in the observable
universe, so that everything would disappear. The doomsday seems
thus unavoidable in this case. However, there could have a way out
from this catastrophe.

Actually, the solution described by Eqs. (2.6) and (2.10) show two
branches around the critical time $t_*$, one along which the
expansion of the universe dramatically accelerates towards the
singularity at $t_*$, and the other for $t>t_*$ that describes a
universe which exponentially decelerates towards zero size as
$t\rightarrow\infty$. A possibility to save any animated or
in-animated structure is therefore assuming the natural or
technological embarking, at the "last minute" before the final
catastrophe, of all these structures in wormholes and ringholes
(grown up by the accelerated expansion to macroscopic sizes from
the submicroscopic constructs that originally pervaded the
gravitational vacuum) with one mouths opening to the expanding
branch and the other opening to the later contracting branch. Note
that since one of the hole mouths would be expanding and the other
mouth would be contracting, every of such holes behaved like
though it were a true time machine, so that the animated or
in-animated beings embarking these holes before the big rip would
travel into their future. The big rip would then be short-cut and
the travelers suddenly found themselves evolving in a contracting
universe (Fig. 2) where all parts that went beyond their horizon -
possibly galaxies, clusters, etc- as they approached the big rip,
should successively re-appear as time progressed in the future.

At least three conditions are needed in order for such a travel to
be feasible: (1) the holes must be enlarged enough to allow
travelers to fit in them, (2) the travelers should not find a
remarkable quantity of exotic matter while traversing the hole,
and (3) the value of the scale factor at the embarking time on the
expanding branch should be the same as the value of that factor at
the arrival time on the contracting branch. Condition (1) can be
fulfilled if the travel takes place at a time sufficiently close
to the big rip which, at the same time, be early enough to keep
the travelers and their surrounding structures reasonably intact.
At the end of Sec. IV it was seen that, at least for a given type
of holes, the exotic character of the hole matter becomes
sufficiently weakened as to allow condition (2) to be satisfied.
On the other hand, Visser, Kar and Dadhich have quite recently
shown [25] that there exist spacetime geometries containing
traversable wormholes which are supported by arbitrarily small
quantities of exotic matter. Finally and quite independent of
these features, if we choose ringholes to play the role of the
time machine, then there are specific itineraries through the
tunneling along which the traveler does not find any exotic matter
[4].

For the case of wormholes in a universe filled with quintessential
dark energy plus a positive cosmological constant, condition (3)
would require the embarking (-) and arrival (+) times to be given
by
\begin{eqnarray}
&&t_{\pm}=\frac{1}{2C_1(|\omega|-1)}\times\nonumber\\
&&\ln\left[\frac{1}{C} +\frac{B}{2a_1^{3(|\omega|-
1)}C^2}\left(1\pm \sqrt{1+\frac{4C a_1^{3(|\omega|-
1)}}{B}}\right)\right] ,\nonumber
\end{eqnarray}
\begin{equation}
\end{equation}
where $a_1$ is any finite sufficiently large value of the scale
factor,
\begin{equation}
C_1=3(|\omega|-1)\sqrt{\lambda}/2
\end{equation}
and
\begin{equation}
B=A/(4C\lambda) .
\end{equation}
We have then
\begin{equation}
t_- < t_* < t_+ .
\end{equation}
Fig. 2 pictorially shows the itinerary followed by the content of
the universe which shortcuts the big rip and a region around it
through a ringhole. It has been recently suggested [26] that there
could be some models for phantom dark-energy for which no big rip
takes place. The discussion included in this section opens the
possibility that even though such models were shown to be
incorrect and phantom energy then necessarily implied a big rip,
the universe could still have in store gravitational procedures
that allowed its inhabitants, and actually its whole content to
circumvent the big rip, while conquering an additional infinite
period of evolution where some laws of physics might be reversed.

\begin{figure}
\includegraphics[width=.9\columnwidth]{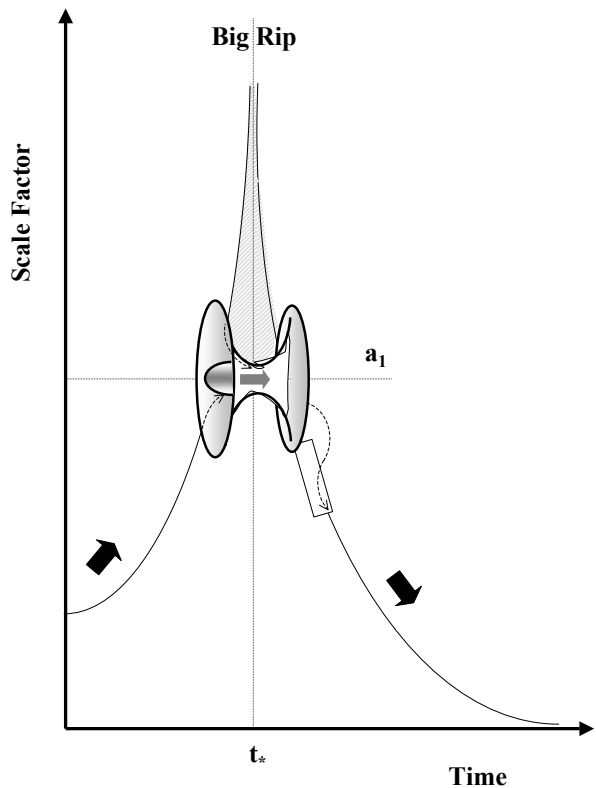}
\caption{\label{fig:epsart} Pictorial representation of the
itinerary followed by a traveler who evolves in an accelerating
universe with phantom energy, avoiding the big rip through a
shortcut made of a ringhole.}
\end{figure}

\section{summary and conclusions}

In this paper we have dealt with the two simplest topologically
different four-dimensional generalizations of Misner space when
embedded in an accelerating universe filled with dark energy
satisfying a general perfect-fluid equation of state,
$p=\omega\rho$, where $-0.8\geq\omega\geq-1.5$, according to the
most recent constraints [16]. We investigate how the overall size
and precise shape of wormholes and rinholes are in this way
induced to evolve due to the rapid expansion of the universe. This
study is carried out adding to the usual quintessence model with
constant $\omega>-1$ the phantom-energy regime where $\omega<-1$
which is not excluded by present observations. It is obtained that
whereas the shape of the holes is always preserved, their size
increases with time relative to the initial embedding-space
coordinate system. With respect to that system, the exoticity
function is also seen to vary with time, though it generally does
not vanish anywhere, even when a big rip model is considered. The
above results allow us to tentatively consider a scenario where,
even though the big rip is taken to be unavoidable for
$\omega<-1$, possible future civilizations might still escape from
the doomsday toward their future by using sufficiently grown up
wormholes and ringholes as time machines. In such a scenario the
suggested difficulty for constructing fundamental theories based
on defining an S matrix could be also circumvented as there will
be no future event horizon for the contracting branch of
cosmological evolution that follows the big rip.

\acknowledgements

\noindent The author thanks Mariam Bouhmadi and Carmen L. Sig\"{u}enza
for useful discussions. This work was supported by DGICYT under
Research Project BMF2002-03758.


\begin{references}
\bibitem {23} A definition of the exoticity function can be found
in Refs. [3] and [21].
\end{references}
\end{document}